\def\a{\alpha}\def\b{\beta}\def\g{\gamma}\def\d{\delta}
\def\r{\rho}\def\ra{\rho_\alpha}
\def\Ea{E_\a}
\def\pia{\pi_\a}\def\pib{\pi_\b}\def\pig{\pi_\g}
\def\Wab{W_{\a\b}}
\documentstyle[11pt]{article}
\begin{document}
\begin{titlepage}
\title{Comments on Omn\`es' Model for Uniqueness of Data}
\author{Lajos Di\'osi\thanks{E-mail: diosi@rmki.kfki.hu}\\
KFKI Research Institute for Particle and Nuclear Physics\\
H-1525 Budapest 114, POB 49, Hungary\\\\
{\it bulletin board ref.: gr-qc/9404064\hfill}}
\maketitle
\begin{abstract}
Standard methods of the theory of permanent state
reduction are shown to offer an alternative realization of Omn\`es' project.
Our proposal,  as simple as Omn\`es' one,
possesses closed master equation for the ensemble density operator,
assuring causality.
\end{abstract}
\end{titlepage}

In his recent Letter \cite{Omnes},
Omn\`es has outlined an appealing concept to generate
unique datum from quantum mechanics modified by a conjectured interaction
between space(-time) and the dynamic system evolving in it.
A concrete stochastic model has been presented.
In our Comment we would like invoke recent ideas,
see Ref.~\cite{Dio89} and references therein, promoting the concept very
much like Omn\`es one. The corresponding theory is a realistic candidate
to solve the data uniqueness problem \cite{Bel90}. It exploits the
theory of permanent state reduction which has emerged from a great
deal of parallel efforts (with milstones such as, e.g., Refs.~[4-12]).
%\cite{Pea76,Pea86,Pea89,Gis84,Gis84Mas,Gis89,Dio88,Dio88Ito,BarBel91}).
These efforts has recently led to {\it standard}
equations of permanent state reduction, i.e. the
{\it quantum state diffusion} theory
\cite{GisPer9294}, extending earlier results
\cite{Dio88} onto arbitrary dimensions.
All these well developed antecedents
invite us to revise (also to correct, in some sense)
the model \cite{Omnes} of Omn\`es.

Omn\`es starts with the {\it strong consistency condition}.
It holds for the
quantum state $\r$ of a macroscopic system if there exists a certain
complete and orthogonal set of Hermitian projectors $\{\Ea\}$ such that
\begin{equation}
\r=\sum_\a\Ea\r\Ea=\sum_\alpha \pia\ra~~~,
\end{equation}
where $\pia=tr\left(\Ea\r\right)$ and $\ra=\pia^{-1}\Ea\r\Ea$.
Initially, say at t=0,
the strong consistency condition (1) may not be satisfied.
As time goes on, {\it decoherence} can successively enforce the approximate
validity of (1). In Omn\`es' model, a conjectured space interaction
on the probability parameters $\pia$ assures decoherence. The $\pia$'s
perform a specially chosen Brownian motion: one $\pia(t)$
will end up becoming $1$ with probability $\pia(0)$;
the other ones will be zero. In such a way yields the model the
uniqueness of data concerning the properties $\{\Ea\}$.

One can (and have to, as we shall argue later)
enrich Omn\`es' work by assuming a simple master equation for
the density operator \cite{Gis84,Pea86,Dio88},
assuring the approximate fulfillment of the
consistency condition (1):
\begin{equation}
\dot\r={\cal L}_0\r-{1\over\tau}\r+{1\over\tau}\sum_\a \Ea\r\Ea
\end{equation}
where ${\cal L}_0$ is the linear evolution superoperator of the system
itself while the further terms on the RHS come from the conjectured
interaction with the space. These terms tend to make $\r$ block-
diagonal on a time scale $\tau$. It will really do it approximately,
against the self-dynamics ${\cal L}_0\r$  which might usually restore
the damped off-diagonals.

Closely related to the master Eq.~(2), let us introduce the
following diffusion matrix \cite{Dio88}:
\begin{equation}
\Wab={1\over\tau}\pia\pib\left(\d_{\a\b}-\pia-\pib+\sum_\g\pig^2\right)~~~.
\end{equation}
Observe that the trace
$w\equiv\sum_\a W_{\a\a}=\tau^{-1}\left(1-\sum_\a\pia^2\right)$ vanishes
{\it iff}
all $\pia$ is zero but one equals to $1$. So, $w$ is a good quantity to
qualify the non-uniqueness of data in question.
Let us replace Omn\`es' diffusion matrix in his Eq.~(4) \cite{Omnes}
by our one (3):
\begin{equation}
<\dot\pia(t)\dot\pib(t^\prime)>=2\Wab\d(t-t^\prime)
\end{equation}
for all $\a,\b$.
For times $t>>\tau$,
the above Brownian motion drives a given $\pia(t)$ to $1$
with probability $\pia(0)$; the other ones tend to $0$.
(see proof, e.g., in Ref.~\cite{Pea86}).
At this level, our model is equivalent
to Omn\`es' one in offering data uniqueness.

What else can our alternative model offer? Most importantly,
closed evolution equation, modified by the concejtured interaction with
space, can be constructed for the system's density operator.
There are separate paths $\r(t)$ for each realization of
the $\pia$'s corresponding to a given pattern of
interaction with space. The corresponding
paths $\r(t)$ are random (Brownian) paths embedded in the space of
density operators. To specify such a $\r$-valued Brownian motion,
let us define the diffusion super-matrix and the drift, too, as follows:
\begin{eqnarray}
& &<\dot\r\otimes\dot\r>= \nonumber\\
        {1\over\tau}\sum_\a
& &\Bigl(\bigl(\Ea-\pia\bigr)\r\otimes\r\bigl(\Ea-\pia\bigr)
        +\r\bigl(\Ea-\pia\bigr)\otimes\bigl(\Ea-\pia\bigr)\r \Bigr)~~, \\
& &<\dot\r>={\cal L}_0<\r>-{1\over\tau}<\r>+{1\over\tau}\sum_\a\Ea<\r>\Ea~~.
\end{eqnarray}
The above equations need a comment each. The diffusion Eq.~(5)
leads directly to the diffusion Eq.~(4) of the probability parameters,
via the relations $\pia=tr\left(\Ea\r\right)$.
The drift term is, as it should be,
identical to the RHS of the master Eq.~(2), apart from the notational
difference. [In Eqs.~(1) and (2), $\r$ denotes the ensemble state;
in the subsequent part, however, the same symbol $\r$ is to denote
the state of a sub-ensemble
of a particular interaction pattern, and $<\r>$ must have been introduced
for the ensemble state.]

What we have presented so far is an alternative concrete realization
of Omn\`es' concept of data uniqueness from modified quantum mechanics.
Due to the achievements of previous parallel researches, perhaps our model
goes beyond Omn\`es' one. In Omn\`es' model no closed evolution
(master) equation exists for the ensemble density operator. This
would lead to acausal effects \cite{Gis84Mas,Gis89}.
Obviously, only models {\it without} the master equation allow
complete reduction within finite time.
Models {\it with} master
equation have asymptotic reduction, not a high price for causality.

The concept of Ref.~\cite{Omnes}
has a further delicate requirement: only the probabilities
$\pia$ of the {\it collective spatial properties} $\Ea$ are to be modified
(in favor of the uniqueness of the latters); the internal quantum
degrees of freedom must behave completely unchanged.
This criterium has been
perfectly met in Ref.~\cite{Dio89}, with a suitable cutoff \cite{GGP90}.
The mechanism, however, differs from that of Omn\'es model.
Let us outline it, changing the original self-consistent presentation and
adopting again the terminology and  the  setting up of Ref.~\cite{Omnes}.

The collective spatial property $\a$
is identified with the mass density
distribution $f$ of the macroscopic system. That $f$ is not countable needs
extra considerations, of course.
The space interaction
is derived from the Newtonian limit of very tiny stochastic
fluctuations of space-time metric, calculated heuristically
(concejtured, after all) \cite{DioLuk8789}.
Then the analogue of the master Eq.~(2) is derived routinely \cite{Dio87}.
{}From the master equation,  the analogues of
diffusion Eqs.~(4-6) follow automatically, according to the quantum state
diffusion theory. As a result of diffusion, the probability parameters
$\pi_f$ of
large scale mass distribution $f$ of the macroscopic system become
unique in the very sense of Omn\`es' concept. At the same time,
the {\it tiny} space fluctuations we started with
will not cause any observable
effect to the microscopic quantum degrees of freedom.

Finally, we risk a filological remark \cite{Dio92}.
For recent years, two
independent schools of succesful researches have been approaching
the same robust problem in quantum theory:
schools of {\it decoherent history} and of {\it quantum state diffusion},
respectively.
Omn\`es' Letter presents a particular example to put the two together.
Our Comment tried, above all, to show that the overlap of the two
is more fertile than thought so far.
A conceptual comparison and unification
of both is under publication \cite{Dioetal}.

\bigskip
This work was supported by the grant OTKA No. 1822/1991.

\end{document}